**Title**
Improved-Mask R-CNN: Towards an Accurate Generic MSK MRI instance segmentation platform (Data from the Osteoarthritis Initiative)

**Authors**
Banafshe Felfeliyan[1,2], Abhilash Hareendranathan[3], Gregor Kuntze[2], Jacob L. Jaremko[3,4], Janet L. Ronsky [1,2,5]

**Affiliations**
[1] Schulich School of Engineering, University of Calgary
[2] McCaig Institute for Bone and Joint Health University of Calgary, Calgary
[3] Department of Radiology & Diagnostic Imaging, University of Alberta
[4] Alberta Machine Intelligence Institute (AMII), University of Alberta
[5] Mechanical and Manufacturing Engineering, University of Calgary, Calgary

**Corresponding author**
Banafshe Felfeliyan

**Present address**
Email: banafshe.felfeliyan@ucalgary.ca
Address: 2500 University Drive N.W., Calgary, Alberta, Canada, T2N 1N4

# 1 Abstract

Introduction:
Objective assessment of osteoarthritis (OA) Magnetic Resonance Imaging (MRI) scans can address the limitations of the current OA assessment approaches. Detecting and extracting bone, cartilage, and joint fluid is a necessary component for the objective assessment of OA, which helps to quantify tissue characteristics such as volume and thickness. Many algorithms, based on Artificial Intelligence (AI), have been proposed over recent years for segmenting bone and soft tissues. Most of these segmentation methods suffer from the class imbalance problem, can't differentiate between the same anatomic structure, or do not support segmenting different rang of tissue sizes. Mask R-CNN is an instance segmentation framework, meaning it segments and distinct each object of interest like different anatomical structures (e.g. bone and cartilage) using a single model. In this study, the Mask R-CNN architecture was deployed to address the need for a segmentation method that is applicable to use for different tissue scales, pathologies, and MRI sequences associated with OA, without having a problem with imbalanced classes. In addition, we modified the Mask R-CNN to improve segmentation accuracy around instance edges.

Methods:
A total of 500 adult knee MRI scans from the publicly available Osteoarthritis Initiative (OAI), and 97 hip MRI scans from adults with symptomatic hip OA, evaluated by two readers, were used for training and validating the network. Three specific modifications to Mask R-CNN yielded the improved-Mask R-CNN (iMaskRCNN): an additional ROIAligned block, an extra decoder block in the segmentation header, and connecting them using a skip connection. The results were evaluated using Hausdorff distance, dice score for bone and cartilage segmentation, and differences in detected volume, dice score, and coefficients of variation (CoV) for effusion segmentation.

Results:
The iMaskRCNN led to improved bone and cartilage segmentation compared to Mask RCNN as indicated with the increase in dice score from 95% to 98% for the femur, 95% to 97% for the tibia, 71% to 80% for the femoral cartilage, and 81% to 82% for the tibial cartilage. For the effusion detection, the dice score improved with iMaskRCNN 72% versus Mask R-CNN 71%. The CoV values for effusion detection between Reader1 and Mask R-CNN (0.33), Reader1 and iMaskRCNN (0.34), Reader2 and Mask R-CNN (0.22), Reader2 and iMaskRCNN (0.29) are close to CoV between two readers (0.21), indicating a high agreement between the human readers and both Mask R-CNN and iMaskRCNN.

Conclusion:
Mask R-CNN and iMaskRCNN can reliably and simultaneously extract different scale articular tissues involved in OA, forming the foundation for automated assessment of OA. The iMaskRCNN results show that the modification improved the network performance around the edges.

# 1 Introduction

Osteoarthritis (OA) is a major source of pain (Barron, Melanie C., 2007), disability (Felson et al., 1987), reduced quality of life (Salaffi et al., 2005) and places a substantial financial burden on the healthcare system (Wright et al., 2010). OA mainly affects weight-bearing joints like knee and hip joints (Sharif et al., 2017). It's prevalence in Canada is high with 1 out of 8 adults suffering from hip and knee OA (Sharif et al., 2017). Traditionally, OA severity is graded through radiographic assessment using the Kellgren-Lawrence (K-L) scale (John Ball; Maurice Rutherford Jeffrey; Jonas Henrik Kellgren;, 1963). However, the K-L grading measure is insensitive to potentially treatable early changes before structural damage is established (Bruyere et al., 2007). Evaluation of the early morphological changes of OA can be achieved through the use of Magnetic Resonance Imaging (MRI) (Emery et al., 2019). Unlike X-ray, MRI can reveal soft tissue, which is useful in the initial stages as OA becomes manifest as degeneration in soft tissues (Choi and Gold, 2011).

To assess the presence and severity of OA features in knee and hip MRI, several scoring systems have been developed. These scoring systems, which assess features like cartilage and meniscus damage, synovitis, and bone marrow lesions include: the MRI OA Knee Score (MOAKS) (Hunter et al., 2011), the Hip OA MRI Score (HOAMS) (Roemer et al., 2011), the Knee Inflammation MRI Scoring System (KIMRISS) (Jacob L Jaremko et al., 2017), and the Whole Organ MRI Score (WORMS) (Peterfy et al., 2004). In most of these MRI scoring systems, pathologies are traditionally assessed holistically as "mild, moderate, or severe", as seen in the HOAMS scoring system (effusion grades 0-3) (Choi and Gold, 2011). Although, these scoring methods have been used in a variety of research studies and have been helpful in better understanding OA over traditional KL grading, they are still subjective, qualitative, not continuous, and insensitive to small changes. Even with more granular scoring systems like KIMRISS (Jacob L Jaremko et al., 2017) or HIMRISS (Deseyne et al., 2018) the scoring of each subregion remains manual and subjective. Quantitative and objective assessments can address the limitations of these semi-quantitative scoring systems. Alizai et al. compared a quantitative cartilage lesion scoring method with WORMS and demonstrated higher accuracy with the quantitative score in detecting changes in cartilage lesions over time (Alizai et al., 2014). Moreover, as new OA treatment options are targeted towards inflammation (Jacob L. Jaremko et al., 2017), with quantitative measurement of effusion volume as a marker of active disease, the OA assessment capability can be magnified by improving sensitivity to changes with time or treatment.

Despite numerous benefits of having quantitative measurements, they have not been adopted widely in the clinical setting. A key barrier to adoption is the pipeline processes for quantitative measurements usually involve segmentation, registration, and abnormality detection. Until recently, these processes were performed manually, which was laborious, slow, and rater dependent. Therefore, automated approaches are needed to enable routine clinical use of quantitative measurements.

## 1.1 Related work

In recent years, progress has been made toward automated OA assessment (Astuto et al., 2021)(Jaremko et al., 2021). Volume quantification of hip effusion has been assessed using interactive semi-automated segmentation techniques to obtain volume quantification measures (Quinn-Laurin et al., 2021). The attempts for detection and segmentation of OA involved tissues (e.g. bone, cartilage, meniscus) and automating scoring methods have been more successful with the advancement of deep learning in medical image processing (Astuto et al., 2021; Dreizin et al., 2019; Liu et al., 2018; Perry et al., 2019; Prasoon et al., 2013; Shah et al., 2019). Astuto et al. (Astuto et al., 2021) showed an automated deep learning grading system constructed from a cartilage compartment segmentation block and three classification blocks can be used by radiologists to aid in making OA assessment less subjective and increase intergrader agreement.

The patch-based multi-planar cartilage segmentation method proposed by Prasoon et al. (Prasoon et al., 2013) was one of the early attempts to use deep learning for musculoskeletal (MSK) segmentation. Prasoon et al. used a Convolutional Neural Networks (CNN) which despite conventional machine-learning methods didn't require feature selection and reported Dice score 82% which approximately was improved 2% in compare previous state-of-the-art methods. Later Liu et al. (Liu et al., 2017) reported accuracy score 62% for a slice-based segmentation of lower limb MR images using the Segnet CNN architecture followed by 3D deformable model post-processing. Shah et al. (Shah et al., 2019) used a deep learning algorithm for cartilage segmentation and thickness calculation. In their study, a deep learning model was trained over 167 MRI scans for segmentation. The cartilage thickness was assessed at four specified points at the cartilage surface by measuring the length of the vector from the point of interest to the femur perpendicular to the cartilage surface. Applying a multivariate linear regression on cartilage thickness in each point revealed that sex and BMI affect average cartilage thickness (Shah et al., 2019) .

Hodgson et. al, (Perry et al., 2019) proposed a method for semi-automatic measurement of synovial volume that used an active appearance modeling algorithm to register a 3D mask to the image for specifying the region of interest (ROI). The mask was obtained with a manually segmented training set. A two-stage segmentation for extracting traumatic pelvic fluid proposed by Dreizin et al. (Dreizin et al., 2019) was successful in obtaining relatively accurate and robust results with Dice score of 71%. The first network identified coarse information and generated an ROI bounding box. The second network provided for fine segmentation, using the results of the previous network to achieve a dense segmentation (Dreizin et al., 2019).

Deep learning methods have enabled considerable improvement in the accuracy of tissue segmentation and feature extraction. However, additional supervision and post-processing steps, such as information from statistical shape modeling, have been employed to refine results of the deep learning methods (Ambellan et al., 2019; Liu et al., 2018). Additionally, there remains a need for a generalized and accurate methodology for segmenting and detecting the various tissue size and abnormalities involved in OA. Moreover, it is challenging to generalize deep learning to MRI images of MSK structures because they contain information from tissues and pathologies with different scales. The shape and size of pathologies vary in different scan slices, e.g., a tiny sub-millimeter cartilage defect vs. a large joint bone. For this reason, it is necessary to deploy a method that can handle the variation of pathology scales.

U-net based approaches (encoder-decoder networks) can handle multi-scale image information due to usage of "skip connections" in different convolutional layers (Felfeliyan et al., 2019; Liu et al., 2018; Ronneberger et al., 2015). A major limitation of the U-net based approaches is that they suffer from the class imbalance problem and perform semantic (or pixelwise) segmentation, which does not differentiate between similar tissues in different locations (e.g. femur and tibia bone tissues). This limitation is addressed by the Mask R-CNN (He et al., 2017) which is one of the most successful instance segmentation models.

Instance segmentation models can distinguish individual instances meaning they are capable of simultaneous segmentation and classification of different anatomical structures (e.g. femur, and cartilage) and multiple effusion regions with different scales using a single model (He et al., 2017)(Jaremko et al., 2021).

Mask R-CNN provides robustness against the class imbalance problem, and can segment many instances simultaneously by providing box-level localization information for instance-level segmentation. This algorithm is based on region proposal-based object detection and is an extended version of Fast R-CNN (Girshick, 2015) and Faster R-CNN (Ren et al., 2017). Mask R-CNN has been successfully used to infer different pathologies in medical image applications like tumor and nodule detection in PET and ultrasound images (Abdolali et al., 2020; Chiao et al., 2019; Jaremko et al., 2021; Zhang et al., 2019) Mask R-CNN has also been successfully trained to localize and segment knee menisci and identify meniscal tears (Couteaux et al., 2019).

Despite the successful performance of Mask R-CNN in detection tasks, challenges persist in performing accurate pixel classification near boundaries of the classified objects, especially for larger objects (Figure 1). Therefore, to address shortcomings of previous deep learning segmentation algorithms of MSK tissues, including the class imbalance problem, the weakness for multiscale segmentation, and lack of ability to differentiate between objects within the same class the objective of this research was to improve the Mask R-CNN deep neural network.

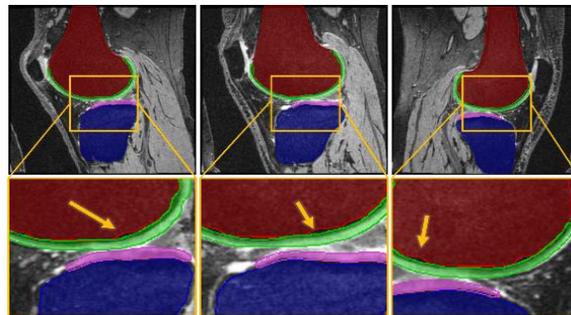

Figure 1: Mask R-CNN performs poorly on the boundary of objects. Although the segmentation accuracy for cartilage (green) appears relatively high on images viewed at a distance, zooming in reveals repeated errors in the bone-cartilage interface (arrows) which are non-physiologic and could mask the small changes in cartilage that are crucial to assessing OA progression.

## 1.2 Contributions

In this paper we developed an "*improved Mask R-CNN*" (iMaskRCNN), by adding skip connections and developed a new technique to consolidate the information from feature maps at multiple scales with high accuracy. The performance of the iMaskRCNN using was evaluated using ResNet and DenseNet backbones. We have tested our algorithm on two datasets. For the first dataset we aim to segment bone and cartilage on the knee (femur (F), tibia (T), femoral cartilage (FC) and tibial cartilage (TC)). For the second dataset we aim to detect effusion of the hip. The iMaskRCNN framework increases the accuracy of the segmentation along the boundaries of the anatomical structures, which could add clinical value in OA assessment.

## 2 Materials and methods

### 2.1 MRI datasets

The segmentation algorithm was developed and evaluated using data from two datasets to examine the segmentation accuracy over different cohorts, sequences, and scanners.

**Osteoarthritis Initiative (OAI):** 500 scans of adult knees from the publicly available OAI dataset were used for bone and cartilage segmentation tasks. Bone and cartilage in the high-resolution sequence of the OAI dataset was previously manually segmented by experts at the Zuse Institute Berlin (ZIB dataset) (Ambellan et al., 2019) (femur, tibia, femoral cartilage, and tibial cartilage pixelwise label available for all slices in the data). The ZIB-OAI dataset contains data from all ranges of OA grades, and more than half of the data are from subjects with severe OA (Ambellan et al., 2019) (Table 1).

Table 1: OAI ZIB summary (Ambellan et al., 2019)

| | |
|---|---|
| MRI scanner | Siemens 3T Trio |
| MRI sequence | DESS |
| Acquisition plane | sagittal |
| Image resolution [mm] | $0.36 \times 0.36 \times 0.7$ |
| Manual segmentations | bones and cartilage |
| Number of subjects | 507 |
| Sex (male, female) | (262,245) |
| Age [years] | $61.87 \pm 9.33$ |
| BMI [kg/m$^2$] | $29.27 \pm 4.52$ |
| rOA grade (0,1,2,3,4) | (60,77,61,151,158) |
| timepoints | baseline |

**Clinical Hip (Effusion) Dataset:** The existing University of Alberta Steroid Injection in Hip Osteoarthritis (STIHO) cohort includes 97 adults with symptomatic hip OA who presented to a radiology clinic for fluoroscopically guided steroid injection. With study ethics approval (UofA HREB Pro00039139) and written informed consent, subjects underwent MRI of both hips' pre-injection and 8 weeks post-injection. Of 97 enrolled patients, no images were available for 4, only 1 time point was available for 6, and only one hip that could be analyzed for 1, leaving 180 hip image sets in 93 patients (Table 2). 175 scans were assessed in random order blinded to chronology and clinical data. Effusion labels were produced semi-automatically by two trained musculoskeletal radiologists using an interactive image segmentation software developed inhouse using MATLAB (MathWorks, USA) software (Quinn-Laurin et al., 2021). The radiologist outlined the ROI around the hip. The software identifies and counts voxels above a threshold determined using Otsu thresholding (Otsu, 1979). The joint fluid volume is calculated as the sum of all identified voxels across all MRI slices.

Table 2: STIHO summary

| | |
|---|---|
| MRI scanner | 3T SIEMENS scanner |
| MRI sequence | Short-TI Inversion Recovery (STIR) |
| Acquisition plane | Wide field-of-view Coronal |
| Body part examined | Both hips |
| Image resolution [mm] | $0.91 \times 0.91 \times 3$ |
| Manual segmentations | Effusion |
| Number of subjects | 93 |
| Number of scans | 180 |
| Sex (male, female) | (51,42) |
| Age [years] | $59 \pm 13$ |

## 2.2 AI Architecture

Mask-RCNN is an instance segmentation model which uses 1) a region proposal network (RPN) to recognize objects and locations, 2) a deep encoder neural network model (called the backbone) to generate features, and 3) two heads, a class-head for classification of extracted bounding boxes and a mask-head responsible for extracting pixel-wise masks from the cropped features obtained from extracted bounding boxes (refer Fig 2). Using a hybrid loss function that simultaneously minimizes the error of region proposal, classification, and segmentation components, the Mask R-CNN identifies object instances and regions of interest. However, since none of its components specifically address accuracy of boundary pixels, Mask R-CNN may not precisely capture the anatomical boundaries which is crucial in medical image quantitative feature assessments. In the iMaskRCNN we address this limitation by modifying the mask head block (Figure 2). Various components of the network are described next.

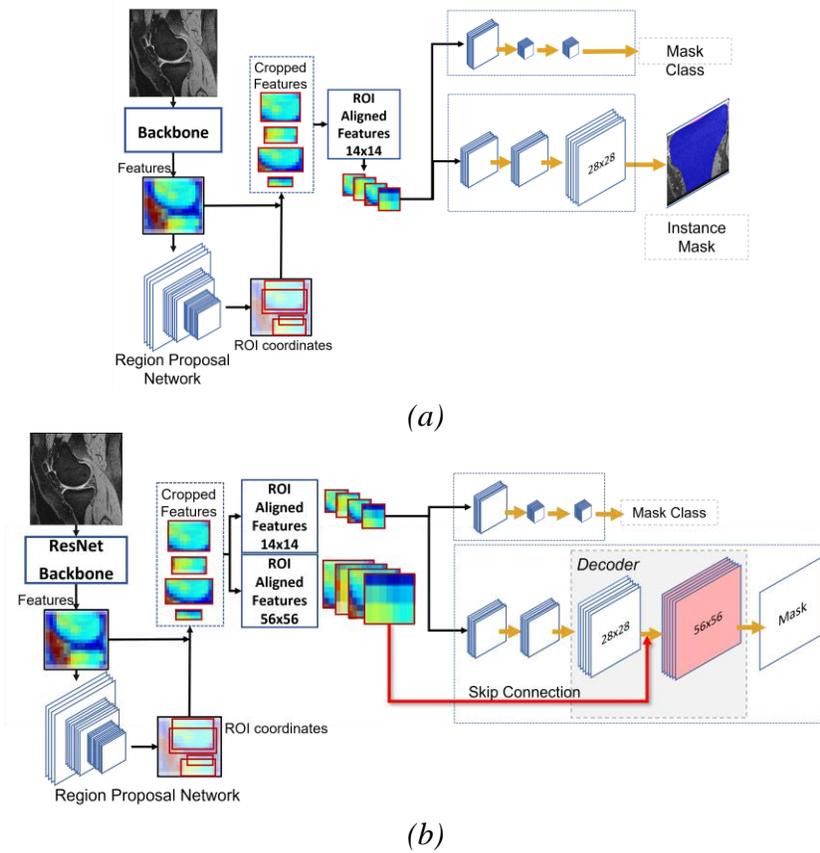

Figure 2: The Pipeline of "Mask R-CNN" (a) and "iMaskRCNN" (b). (b) The decoder block is added to increase features' spatial resolution. Furthermore, the higher resolution features (56x56) obtained from ROIAligned block are concatenated to the decoder to pass spatial information lost during down-sampling.

**Backbone Network:** Two popular deep encoder models – ResNet101-FPN (ResNet (He et al., 2016) + Feature pyramid network (FPN) (Lin et al., 2017)) and DenseNet (Huang et al., 2017) were used as backbone models (**Figure 3**). The Resnet-FPN backbone, deployed in the original Mask R-CNN, extracts feature maps of the input in different scales and resolution. The FPN structure added to the ResNet extracts multiscale semantic features via bottom-up and top-down pathways, and lateral connections (**Figure 3**) (Zhang et al., 2020).

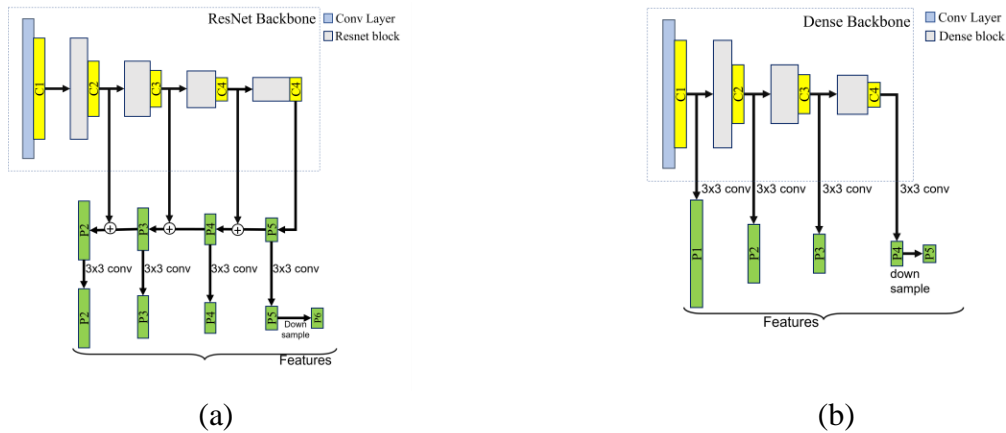

(a) (b)

Figure **3** : (a) The Resnet-FPN Backbone, (b) The Dense Backbone

**Region Proposal Network (RPN):** Features from the backbone are passed into the RPN block that identifies objects of each class (e.g. bone, cartilage, effusion) and calculates the bounding box around the object. Since the RPN receives features from different layers of the backbone it can extract different objects with various scales. Large objects are extracted from low-resolution feature maps (e.g. femur) and smaller objects (e.g. effusion) are extracted from high-resolution feature maps.

**ROIAlign**: The "ROI pooling layer" is the block for obtaining the small feature map of the ROI bounding box used in Fast R-CNN (Girshick, 2015). It resizes ROI features into a small fixed spatial extent feature map using max-pooling. During this pooling process for feature extraction there will be misalignment in segmentation results due to the spatial quantization process. The

*ROIAlign* block (14×14) available in the Mask R-CNN was applied to fix this problem by canceling the rounding operation that happens during ROI resizing process. Further, we proposed to have a second ROIAlign layer (56 × 56) which generates a higher resolution feature that will be utilized by the modified mask head (Figure 2).

**Classification head:** The existing Mask R-CNN classification head was used to perform ROI classification and bounding-box regression.

**Modified Mask head:** Two key modifications to the mask head were introduced, to potentially enable the network to detect anatomical boundaries more precisely. First, a decoder layer was added to the mask head to increase the spatial resolution of feature maps by learnable up-sampling. The transposed convolution was used as the decoder layer, as it is reportedly one of the best learnable up-sampling layers for the instance edge segmentation task (Wojna et al., 2019). Second, the high-resolution features obtained from the second ROIAlign block were fused into the added decoder layer with a skip connection which transports features from the higher layer of the network.

**Loss**: As in the Mask R-CNN, our model has a multi-task learning loss ($L$) for each sampled ROI which is the result of the classification loss, the bounding-box loss, and the mask-loss accumulation (eq. (1)).

$$L = L_{cls} + L_{bbox} + L_{mask} \qquad (1)$$

where $L_{cls} = class\ label\ prediction\ loss$, $L_{bbox} = bounding\ box\ refinement\ loss$, and $L_{mask} = segmentation\ mask\ prediction\ loss$.

The benefit of using a multi-task learning loss such as eq. (1) is that it helps to achieve a better generalization performance (Goodfellow et al., 2016).

### 2.3 Implementation Detail

The Mask R-CNN (He et al., 2020) was used as the baseline and our model was developed using TensorFlow v2. Backbones included the ResNet-101-FPN pre-trained on the COCO 2017 dataset (Microsoft COCO: Common Objects in Context) (Lin et al., 2014) and a DenseNet (Huang et al., 2017). The iMaskRCNN and the original Mask R-CNN were trained and evaluated on two different datasets related to OA (OAI knee OA and STIHO Hip OA). All hyper-parameters remained the same for both the improved and original Mask R-CNN. The input images were resized to 1024×1024. All models were trained on one NVIDIA V100 GPU for 300 epochs using Adam optimizer and a learning rate of 0.001. The training and testing for segmentation of bone and cartilage instances were carried out using 470 and 30 images from the OAI dataset, respectively. For the effusion segmentation, training and testing were performed on 125 and 50 scans from the STIHO dataset, respectively.

### 2.4 Evaluation Metrics

#### 2.4.1 Bone and Cartilage segmentation

Evaluation of the segmentation algorithm on bone and cartilage was performed using precision and the Dice Similarity Coefficient and the Hausdorff distance.

The precision metric is defined as eq. (2).

$$Precision = \frac{TP}{TP + FP} \qquad (2)$$

Where TP= True positive and FP= false positive.

The Dice coefficient quantifies the degree of overlap between the ground truth and segmentation. The Dice value defined as eq. (3) and is between [0,1], higher values reflect higher similarity:

$$Dice\ (DCS) = \frac{2 \times TP}{(TP + FP) + (TP + FN)} \qquad (3)$$

Where TP= True positive, FP= false positive and FN=false negative.

Hausdorff distance (Unter Rote, 1991) was used for distance assessment between the ground truth and automatic segmentation. Hausdorff distance $H(A, B)$ is defined as the maximum Euclidean distance $(d)$ of nearest points in the two sets $(A\ and\ B)$, this measurement is sensitive to outliers and is defined as eq. 4.

$$H(A, B) = max\ \{max\{\min_{a \in A} d(a, b)\}, max\{\min_{b \in B} d(b, a)\}\} \qquad (5)$$

Where $a$ and $b$ are single points and $a \in A$ and $b \in B$.

Smaller H value indicates better performance of the algorithm.

### 2.4.2 Effusion Detection

The performance of the algorithm for the effusion task was also evaluated with the calculated dice and precision values. Volumetric quantitative measurements (VQM) based on fluid volume were conducted to obtain a better understanding of the results from a clinical perspective. To evaluate effusion detection tasks, all voxels detected as hip fluid were counted and the total number of voxels were compared with the number of voxels detected by two radiologists (human readers). For each reader pair, the mean ± standard deviation (SD) of VQM were calculated and the difference between measurements were determined. Additionally, the coefficients of variation (CoV) between each pair were determined.

## 3 Results

### 3.1 Bone and Cartilage

Qualitatively, both Mask R-CNN and iMaskRCNN produce results with high visual agreement with the ground truth overall for bone and cartilage segmentation, as demonstrated for a set of 4 sample images (Figure 4). However, the manual segmentation missed part of the femur (Figure 4, column 4, yellow arrow), while the iMaskRCNN was able to correctly detect this small area. Modifying the mask head block of the Mask R-CNN successfully improved the segmentation near the bone and cartilage edges (Figure 5).

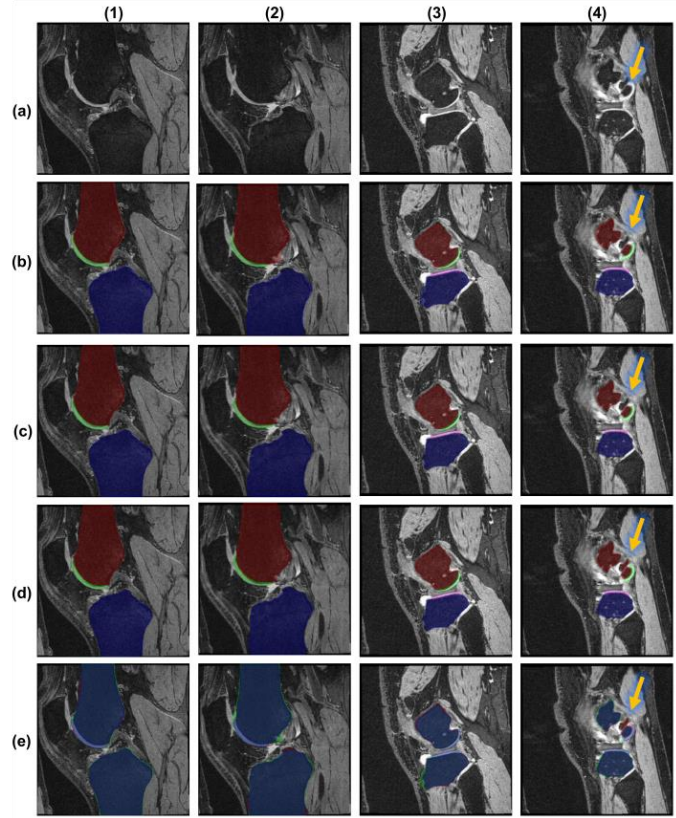

Figure 4: Results of Applying Mask RCNN and iMaskRCNN algorithm to the OAI dataset, (a)Original Image, (b) Ground truth, (c) Mask R-CNN results, (d) iMaskRCNN results, (e) difference between ground truth and result of iMaskRCNN (green = only human reader, red = model detection, Blue = overlap between model and human reader). As demonstrated in row (e) there is a high agreement between the iMaskRCNN results and the ground truth. In column (4), we can observe the human reader has missed part of the femur (row (b) the yellow arrow), while the iMaskRCNN was able to detect this part of the bone (row (d) the yellow arrow). This shows that while the error in manual labeling is inevitable, the network performed well.

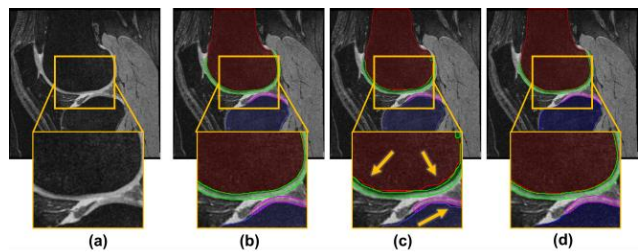

Figure 5: (a) Original Image, (b) Ground truth, (c) Mask R-CNN results, yellow arrows show areas that Mask-RCNN performed poorly on segmentation, (d) iMaskRCNN results.

Quantitative results of the bone and cartilage instance segmentation with the ResNet backbone and DenseNet backbone are presented in Table 3. With the ResNet backbone the iMaskRCNN improved Dice scores significantly (p-value<0.125). The Dice value increased for the F (98±0.0%), T (97±0.0%), FC (86±0.02%) and TC (82±0.04%) compared to the Dice scores of the original Mask R-CNN of F (95±0.0%), T (95±0.0%), FC (71±0.03%) and TC (81±0.03%) (Table 3). The average Hausdorff distance is lower for iMaskRCNN and overall is lower than one pixel (0.06 to 0.3 of a pixel) for both Mask-RCNN and iMaskRCNN. The mean average precision values of the iMaskRCNN (F (95%), T (95%), FC (86%) and TC (88%)) are higher than those of the original Mask R-CNN (F (93%), T (93%), FC (67%) and TC (78%)). Overall, comparing metrics (Table 3) suggests that the iMaskRCNN performs better than the original Mask R-CNN on the segmentation task. The performance of iMaskRCNN vs Mask R-CNN using DenseNet as the backbone network was also evaluated. Similar to results for the ResNet, the iMaskRCNN with DenseNet backbone showed higher performance compared to Mask R-CNN on the OAI knee dataset. For example, Dice values for the iMaskRCNN are F (93%), T (94%), FC (75%), TC (78%) in comparison to the original Mask R-CNN values of F (91%), T (92%), FC (70%), and TC (74%) (Table 3). However, the DenseNet backbone did not show improvements over ResNet (Table 3).

Table 3: Quantitative results of the bone and cartilage instance segmentation

| OAI | Dice | | | | Hausdorff distance (pixel) | | | | Ave. Hausdorff distance | | | | Average precision | | | |
|---|---|---|---|---|---|---|---|---|---|---|---|---|---|---|---|---|
| Tissue | F | T | FC | TC | F | T | FC | TC | F | T | FC | TC | F | FC | T | TC |
| **Mask R-CNN** | 0.95± 0.0 | 0.95± 0.0 | 0.71± 0.03 | 0.81± 0.03 | 6.88± 3.1 | 7.96± 1.4 | 15.2± 5.6 | 14.51 ±1.8 | 0.07± 0.0 | 0.23± 0.0 | 0.47± 0.05 | 0.30± 0.06 | 0.93± 0.0 | 0.93± 0.0 | 0.67± 0.04 | 0.78± 0.07 |
| **iMask RCNN** | **0.98± 0.0** | **0.97± 0.0** | **0.86± 0.02** | **0.82± 0.04** | **6.00± 2.3** | **6.00± 1.5** | **11.9± 3.7** | **13.3± 3.9** | **0.06± 0.0** | **0.10± 0.0** | **0.30± 0.04** | **0.30± 0.05** | **0.95± 0.0** | **0.95± 0.0** | **0.86± 0.03** | **0.88± 0.03** |
| **Mask R-CNN (DenseNet BB)** | 0.91± 0.06 | 0.92± 0.12 | 0.7±0.24 | 0.74± 0.06 | 29.1± 20.3 | 24.0± 25.7 | 22.29 ±14 | 17.9± 11.5 | 2.01± 1.1 | 1.11± 1.3 | 0.90± 2.0 | 0.60± 0.8 | 0.90± 0.03 | 0.91± 0.05 | 0.89± 0.1 | 0.70± 0.08 |
| **iMask RCNN (DenseNet BB)** | 0.93± 0.04 | 0.94± 0.06 | 0.75± 0.8 | 0.78± 04 | 15.01 ±5.7 | 13.40 ±7.6 | 21.2± 14.6 | 16.01 ±3.0 | 0.26± 0.26 | 0.50± 1.0 | 0.50± 0.20 | 0.55± 2.0 | 0.91± 0.3 | 0.94± 0.1 | 0.70± 0.1 | 0.78± 0.2 |

To check whether the segmentation methods are over- or under-estimating, the distance between ground truth and model predictions of tissue surfaces generated by the VTK Python package (visualization toolkit package) (Figure 6) have been calculated (Table 4). As presented in Table 4, since the average of mean distances for both methods is below 0.364 millimeters (=less than one pixel) and close to zero, we can conclude that neither introduces bias.

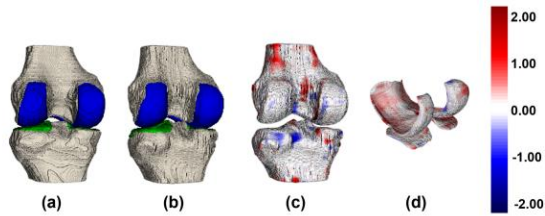

Figure 6: (a) 3D bone and cartilage model generated by GT labels, (b) 3D bone and cartilage model generated by results of IMaskRCNN, (c, d) distance between GT and prediction (mm), red prediction overestimation, blue prediction underestimation, white no difference

Table 4: Surface Distance between tissue labels of ground truth and predictions by two methods

| Tissue Surface | Distance between two surfaces | Average of mean distance (mm) | Average mean distance SD (mm) |
|---|---|---|---|
| FC | $GT - P_{MaskRCNN}$ | 0.13±0.10 | 0.54±0.15 |
| | $GT - P_{iMaskRCNN}$ | **0.03±0.16** | 0.41±0.09 |
| TC | $GT - P_{MaskRCNN}$ | 0.01±0.07 | 0.39±0.11 |
| | $GT - P_{iMaskRCNN}$ | 0.2±0.08 | 0.90±0.12 |
| Femur | $GT - P_{MaskRCNN}$ | -0.35±0.07 | 0.49±0.05 |
| | $GT - P_{iMaskRCNN}$ | **0.16±0.13** | 0.47±0.06 |
| Tibia | $GT - P_{MaskRCNN}$ | 0.47±0.16 | 0.46±0.07 |
| | $GT - P_{iMaskRCNN}$ | **0.05±0.18** | 0.54±0.12 |

Furthermore, the cartilage thickness contained by deep learning models was compared to the cartilage thickness contained by ground truth labels in order to verify the sensitivity of the cartilage segmentation method to cartilage thickness variations. This comparison was conducted by creating cartilage thickness maps of ground truth and predictions, calculating cartilage thicknesses average and standard deviation, and calculation of the cartilage thickness differences.

The 3D cartilage thickness maps were calculated by measuring the distance between the cartilage bony surface and the articular similar to method proposed by Kauffmann et al (Kauffmann et al., 2003). The 2D cartilage thickness map was generated by projecting this distance map onto a 2D plane. The femoral cartilage thickness map was projected radially around the intercondylar axis (Figure 7), The 2D cartilage thickness map was generated by projecting this distance map onto a 2D plane. The femoral cartilage thickness map was projected radially around the intercondylar axis Figure 8 for one subject. As shown in Figure 8, result of both models are very close to the ground truth, however the cartilage thickness of the iMaskRCNN model has less error compared to the result of the original Mask R-CNN. Note that most areas are white indicating very small difference between the prediction and the ground truth.

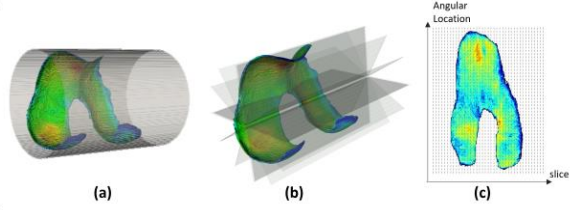

Figure 7: Generating flattened cartilage thickness map: (a) fitting a cylinder around the cartilage, (b) Cartilage thickness map radial projection per degree, (c) femoral cartilage flattened cartilage thickness map plotted on the angular location of the cartilage against the slice number

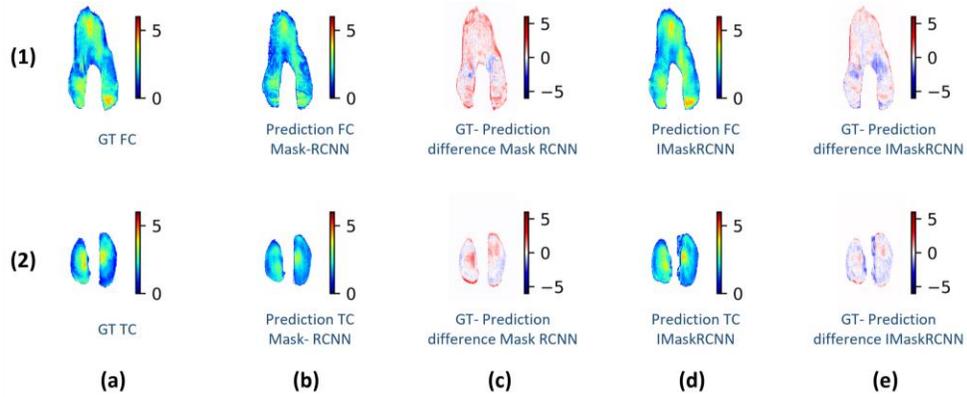

Figure 8: Comparing cartilage thickness map of prediction and ground truth. (Color bar scale is in mm) (a) FC and TC thickness maps Ground truth, (b) FC and TC predicted cartilage maps by Mask R-CNN, (c) difference between (a) and (b) maps, red = under estimation, white = no difference, blue = over estimation, (d) FC and TC predicted cartilage maps by iMaskRCNN, (c) difference between (a) and (d) maps, red = under estimation, white = no difference, blue = over estimation,

Furthermore, the difference between the predictions by Mask R-CNN and iMaskRCNN and the ground truth have been analyzed and represented in Table 5. As shown in the table average-of-means is near to zero, for both methods. However, the absolute value of average-of-means for iMaskRCNN for both cartilages is less than Mask R-CNN showing the modification improved the segmentation results.

Table 5: Difference between ground truth cartilage thickness map and predictions thickness map

| Cartilage | Thickness Difference | Average of mean differences (mm) | Average of SD of mean differences (mm) |
|---|---|---|---|
| FC Thickness | $GT - Mask\ RCNN$ | 0.42±0.13 | 0.72±0.05 |
| | $GT - IMaskRCNN$ | -0.046±0.16 | 0.68±0.05 |
| TC Thickness | $GT - Mask\ RCNN$ | 0.18±0.15 | 0.72±0.08 |
| | $GT - IMaskRCNN$ | -0.25±0.20 | 0.72±0.06 |

## 3.2 Effusion Detection

Applying iMaskRCNN on the STIHO dataset for the effusion detection task, visualized in Figure 9 and Figure 10, resulted in successful major overlap with the ground truth labels.

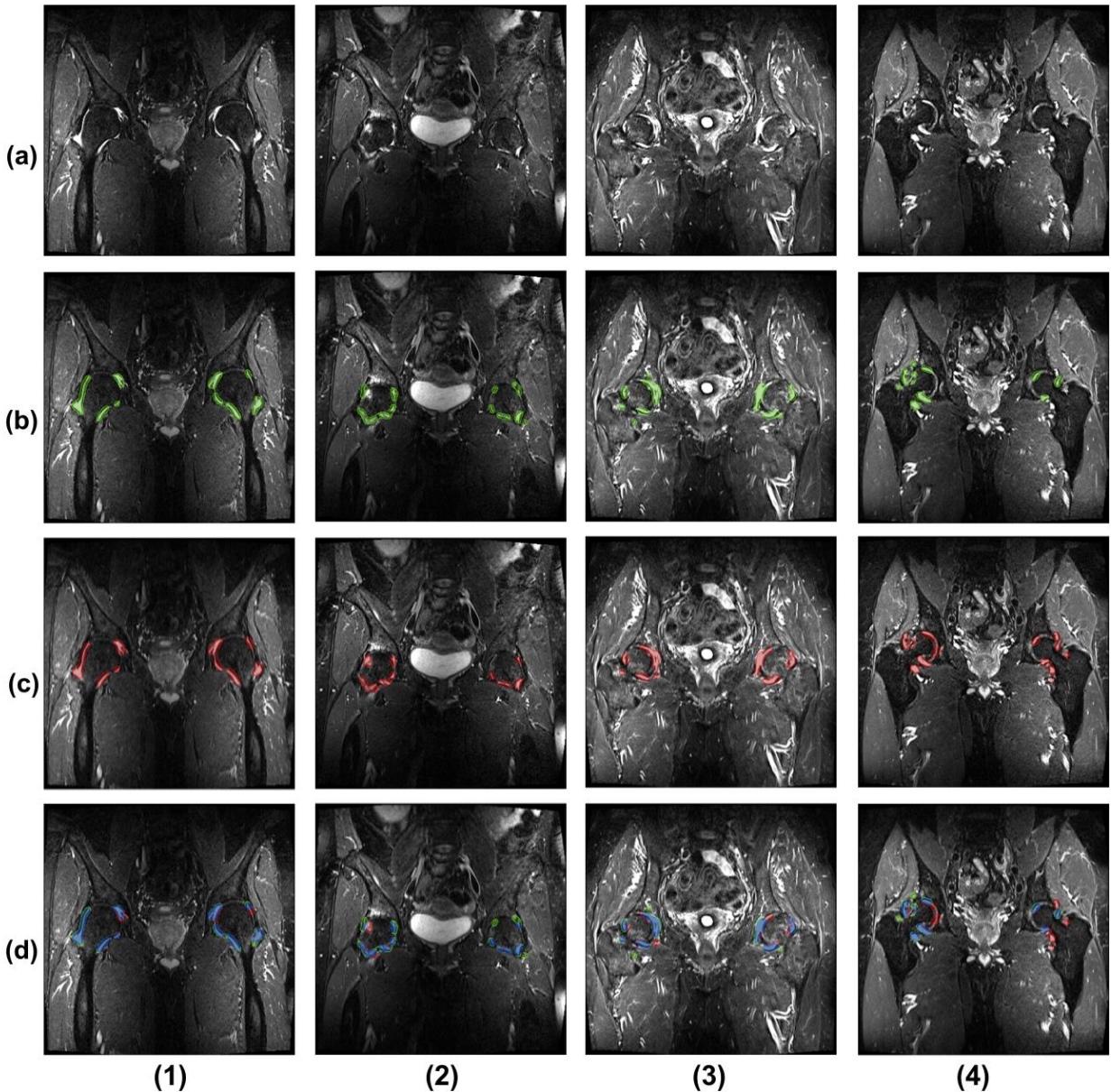

Figure 9: Results of Applying iMaskRCNN algorithm to the STIHO dataset, (a) Original Image, (b) Ground truth, (c) iMaskRCNN results, (d) difference between ground truth and result of I-Mask R-CNN (green = only human reader, red = model detection, Blue = overlap between model and human reader)

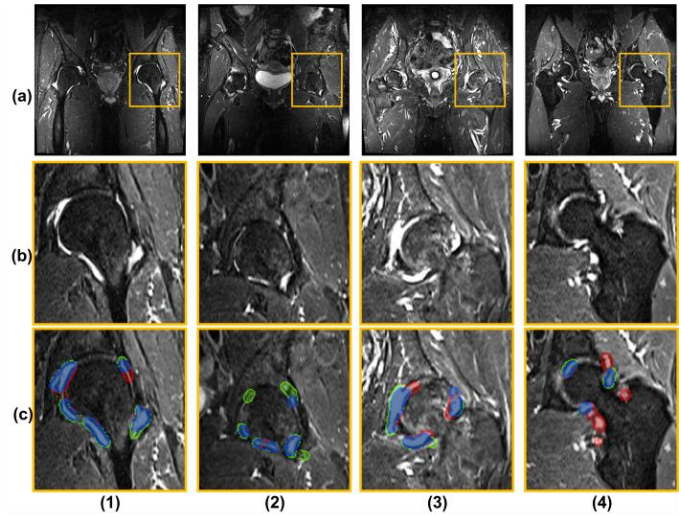

Figure 10: Detail view showing results of applying iMaskRCNN algorithm to the STIHO dataset, (a) Original Image, (b) zoomed in images, (c) difference between ground truth and result of iMaskRCNN (green = only human reader, red = model detection, Blue = overlap between model and human reader)

The Dice values were comparable for the Mask R-CNN and iMaskRCNN, of 0.71 and 0.72, respectively (Table 6).

Table 6: Effusion dice and precision values

| STIHO | Dice | Average precision |
|---|---|---|
| **Mask R-CNN** | 0.71±0.1 | 0.80±0.3 |
| **iMaskRCNN** | **0.72**±0.5 | 0.75±0.1 |

Although a slight difference can be observed visually between the ground truth for effusion and the automatic detection in specific regions (Figure 9 and Figure 10), there is high agreement on the volume (Table 6). The CoV values between the Mask R-CNN and readers (Table 7) were 0.33 (Reader1) and 0.22 (Reader2), and for iMaskRCNN were 0.34 (Reader1) and 0.29 (Reader2). The CoV values between reader2 and the automated methods are lower than reader1, indicating higher agreement between reader2 and automated algorithms. These values of 0.22 and 0.29 are close to the CoV value between the two readers which is 0.21.

Table 7: effusion detected volume comparison

| STIHO | Fluid volumes (Baseline) | | | Difference in volumes | | | | CoV | | | |
|---|---|---|---|---|---|---|---|---|---|---|---|
| | Right | Left | Overall | Reader 1 | Reader 2 | MaskR-CNN | iMaskRCNN | Reader 1 | Reader 2 | MaskR-CNN | iMaskRCNN |
| Reader 1 | 7.72 ± 5.33 | 6.17 ± 6.28 | 6.94 ± 5.85 | 0 | 1.13 ± 0.90 | 3.06 ± 1.77 | 4.22±2.45 | 0 | 0.21 | 0.33 | 0.34 |
| Reader 2 | 8.65 ± 5.05 | 6.63 ± 6.02 | 7.64 ± 5.62 | 1.13 ± 0.90 | 0 | 2.26 ± 1.31 | 3.33 ± 2.04 | 0.21 | 0 | 0.22 | 0.29 |
| Mask R-CNN | 10.05 ± 3.76 | 8.49 ± 3.56 | 9.27 ± 3.73 | 3.06 ± 1.77 | 2.26 ± 1.31 | 0 | NA | 0.33 | 0.22 | 0 | NA |
| iMaskRCNN | 11.81 ± 4.55 | 9.72 ± 4.09 | 10.76 ± 4.43 | 4.22±2.45 | 3.33 ± 2.04 | NA | 0 | 0.34 | 0.29 | NA | 0 |

## 4 Discussion

The objective of this research was to develop and evaluate a method for OA related pathology segmentation and detection on MRI that would be able to accurately segment multiple tissues with different scales at the same time with minimized effects of the class-imbalance problem. Many current segmentation architectures need additional user interaction and post-processing steps (Ambellan et al., 2019; Liu et al., 2018). Furthermore, OA has an unmet need for a generalized method that can accurately detect the various sizes and scales, and abnormalities associated with the disease features. Encoder-decoder networks have partly solved this problem but still suffer from class imbalances and can't perform instance segmentation.

To address these concerns, in this paper the Mask R-CNN algorithm for MSK segmentation was improved and adapted by modifying the 'head network' architecture to improve results at boundaries of anatomical structures. The high-level semantic features are extracted by the backbone and first layers of the mask head. A decoder layer was added combined with the skip connection which passes the high-resolution feature map from the extra ROIAligned network. This approach ensures the recovered feature map integrates low-level features with high level semantic features, which is helpful for segmentation and edge recovery. This simple addition provides small but potentially crucial improvements in modeling of anatomically significant boundaries such as the bone-cartilage interface or cartilage surfaces. The effect of adding skip connections and decoders is well studied (Wojna et al., 2019) and have resulted in higher performance in other semantic segmentation networks (Ronneberger et al., 2015), and supports the idea that adding these blocks would improve instance segmentation and object detection.

The effectiveness of our proposed network was evaluated on knee (OAI dataset) and hip effusion (STIHO) MRI datasets using Dice score, precision and Hausdorff metrics for bone and cartilage segmentation, and Dice and VQM for effusion detection. Furthermore, a comparison of cartilage thickness and surface distance between ground truth and deep learning models was conducted to determine the resiliency of bone and cartilage segmentation.

Despite the presence of a high number of knees with severe OA in the OAI dataset (more than 60% of subjects had radiographic OA grade of 3 or 4) which are generally harder to segment due to large morphological change, the proposed method provided high accuracy in segmentation. This high-performance level is crucial for clinical relevance in these cases that are typically difficult to detect and assess.

Adding the decoder and fusing higher dimensional features lead to a better result by achieving a better edge recovery in larger instances (bones) or instances with high border length (cartilages). However, this approach did not improve the result of segmentation for smaller instances (small effusions).

One potential explanation for this phenomenon is the reshaping and resizing process that the ROI aligned block of Mask R-CNN performs on features. The "ROIAlign" block reshapes ROI cropped features (with different dimensions) obtained from the RPN block into $14 \times 14$ pixels. This reshaping and resizing process is not problematic for small objects and does not impact segmentation accuracy, because there is no (or low) information loss. However, reshaping and resizing larger ROIs into $14 \times 14$ is a downsampling process which causes more information loss which can alter the segmentation result. The original Mask R-CNN network is not able to recover this information during the up-sampling process. For the iMaskRCNN we added another ROIAligned block and features also get reshaped to $56 \times 56$ to enhance detection of abnormalities at a different scale, which as shown in the results was successful.

The ResNet-FPN backbone structure provide excellent multiscale feature extraction for both the iMaskRCNN and Mask R-CNN (He et al., 2017).

To examine the generalizability of modification in the mask head regardless of the utilized backbone, we substituted the original Resnet-FPN with DenseNet for the backbone. The DenseNet network is considered a suitable backbone candidate for Mask R-CNN architecture. In DenseNet, each layer receives inputs from all previous layers and passes its feature maps to the next layers, keeping network parameters low. Furthermore, DenseNet reportedly integrates features from different depths and extracts higher level patterns and produces smoother decision boundaries (Huang et al., 2017). Although the DenseNet backbone did not improve results in comparison to the ResNet-FPN in this current study, the DenseNet iMaskRCNN provided improved results in comparison to DenseNet Mask R-CNN. This finding supports the concept that adding a decoder and skip connection is helpful to obtain more accurate results.

The main challenge faced in this work is that the gold standard used for the training method is human segmentation, which is inevitably subject to variability which would affect the method evaluation procedure. This kind of error is present in most labelled medical datasets. The presence of segmentation errors or noise in the medical image labels by human raters is inevitable and might be attributed to various reasons like rater's fatigue, human error, partial volume effect or subjective labeling. The overall objective for this research is to develop a network robust to label mistakes and with consistent regularization. In the findings from this study (e.g., Figure 4 (col 4 – the yellow arrow) and Figure 10 (col 4)) that the iMaskRCNN performs robustly in the areas that human readers or the semi-automatic algorithm might have incorrectly labelled.

The STIHO labels are generated semi-automatically by human supervision of a thresholding algorithm, which adds some noise to the labels (e.g., sometimes bone marrow edema was selected as the effusion, or some effusions were missed) and consequently degrade quality of label. It can be observed that Mask R-CNN and iMaskRCNN were still able to learn from these weak labels and regularize the detection task.

An additional challenge in the semi-automatically generated STIHO labels, is the variability existing between experts in determining effusion regions in STIHO dataset. This variability reduced the Dice scores.

## 5 Conclusion

OA-related tissues and pathologies are accurately identified and segmented using the proposed iMaskRCNN architecture in this study. This straightforward modification to the original Mask R-CNN has resulted in a significant improvement for clinically important boundaries, especially for larger objects. The impact of the proposed method was observed for both knee and hip which are the two main joints affected by OA. Based on the results obtained, the architecture is concluded to be a good candidate for instance segmentation of different MSK tissues with different sizes. The proposed method is more sensitive to the instance details than the original Mask R-CNN. This accurate segmentation method enables monitoring of changes in OA images and assists in tracking the OA progression or monitoring the effect of medical interventions. Future work to make the evaluation of OA more comprehensive involves adapting this model for bone marrow lesion and knee effusion detection and segmentation.

# 6  Acknowledgments

BF is supported by an Alberta Innovates Graduate Student Scholarship for Data-Enabled Innovation. JJ is supported by a Canada CIFAR AI Chair. Academic time for JJ, RL and VQL is made available by Medical Imaging Consultants (MIC), Edmonton, Canada. We thank the members of the OMERACT MRI in Arthritis Working Group for their participation and support in this project.

# 7  Conflict of interest

There is no conflict of interest, financial or otherwise.

# 8  CRediT authorship contribution statement

**Banafshe Felfeliyan**: Conceptualization, Methodology, Software, Data curation, Data analysis/interpretation, Visualization, Writing - literature research, original draft, review & editing.
**Abhilash Hareendranathan**: Conceptualization, Methodology, Data curation, Data analysis/interpretation, Writing - literature research, review & editing.
**Gregor Kuntze**: Conceptualization, Writing - literature research, review & editing.
**Jacob Jaremko**: Supervision, Funding acquisition, Clinical studies, Writing - review & editing.
**Janet Ronsky**: Supervision, Funding acquisition, Writing - review & editing.

# 9 References

Abdolali, F., Kapur, J., Jaremko, J.L., Noga, M., Hareendranathan, A.R., Punithakumar, K., 2020. Automated thyroid nodule detection from ultrasound imaging using deep convolutional neural networks. Comput. Biol. Med. 122. https://doi.org/10.1016/j.compbiomed.2020.103871

Alizai, H., Virayavanich, W., Joseph, G.B., Nardo, L., Liu, F., Liebl, H., Nevitt, M.C., Lynch, J.A., McCulloch, C.E., Link, T.M., 2014. Cartilage lesion score: Comparison of a quantitative assessment score with established semiquantitative MR scoring systems. Radiology 271, 479–487. https://doi.org/10.1148/radiol.13122056

Ambellan, F., Tack, A., Ehlke, M., Zachow, S., 2019. Automated segmentation of knee bone and cartilage combining statistical shape knowledge and convolutional neural networks: Data from the Osteoarthritis Initiative. Med. Image Anal. 52, 109–118. https://doi.org/10.1016/j.media.2018.11.009

Astuto, B., Flament, I., K. Namiri, N., Shah, R., Bharadwaj, U., M. Link, T., D. Bucknor, M., Pedoia, V., Majumdar, S., 2021. Automatic Deep Learning–assisted Detection and Grading of Abnormalities in Knee MRI Studies. Radiol. Artif. Intell. 3, e200165. https://doi.org/10.1148/ryai.2021200165

Barron, Melanie C., B.R.R., 2007. Managing osteoarthritic knee pain. J. Am. Osteopath. Assoc. 107 Supplement, ES21.

Bruyere, O., Genant, H., Kothari, M., Zaim, S., White, D., Peterfy, C., Burlet, N., Richy, F., Ethgen, D., Montague, T., Dabrowski, C., Reginster, J.Y., 2007. Longitudinal study of magnetic resonance imaging and standard X-rays to assess disease progression in osteoarthritis. Osteoarthr. Cartil. 15, 98–103. https://doi.org/10.1016/j.joca.2006.06.018

Chiao, J.Y., Chen, K.Y., Ken Ying-Kai Liao, Hsieh, P.H., Zhang, G., Huang, T.C., 2019. Detection and classification the breast tumors using mask R-CNN on sonograms. Med. (United States) 98. https://doi.org/10.1097/MD.0000000000015200

Choi, J.A., Gold, G.E., 2011. MR imaging of articular cartilage physiology. Magn. Reson. Imaging Clin. N. Am. https://doi.org/10.1016/j.mric.2011.02.010

Couteaux, V., Si-Mohamed, S., Nempont, O., Lefevre, T., Popoff, A., Pizaine, G., Villain, N., Bloch, I., Cotten, A., Boussel, L., 2019. Automatic knee meniscus tear detection and orientation classification with Mask-RCNN. Diagn. Interv. Imaging 100, 235–242. https://doi.org/10.1016/j.diii.2019.03.002

Deseyne, N., Conrozier, T., Lellouche, H., Maillet, B., Weber, U., Jaremko, J.L., Paschke, J., Epstein, J., Maksymowych, W.P., Loeuille, D., 2018. Hip Inflammation MRI Scoring System ( HIMRISS ) to predict response to hyaluronic acid injection in hip osteoarthritis. Jt. Bone Spine 85, 475–480. https://doi.org/10.1016/j.jbspin.2017.08.004

Dreizin, D., Zhou, Y., Zhang, Y., Tirada, N., Yuille, A.L., 2019. Performance of a Deep Learning Algorithm for Automated Segmentation and Quantification of Traumatic Pelvic Hematomas on CT. J. Digit. Imaging. https://doi.org/10.1007/s10278-019-00207-1

Emery, C.A., Whittaker, J.L., Mahmoudian, A., Lohmander, L.S., Roos, E.M., Bennell, K.L., Toomey, C.M., Reimer, R.A., Thompson, D., Ronsky, J.L., Kuntze, G., Lloyd, D.G., Andriacchi, T., Englund, M., Kraus, V.B., Losina, E., Bierma-Zeinstra, S., Runhaar, J., Peat, G., Luyten, F.P., Snyder-Mackler, L., Risberg, M.A., Mobasheri, A., Guermazi, A., Hunter, D.J., Arden, N.K., 2019. Establishing outcome measures in early knee osteoarthritis. Nat. Rev. Rheumatol. 15, 438–448. https://doi.org/10.1038/s41584-019-0237-

Felfeliyan, B., Kupper, J., Forkert, N., Ronsky, J., 2019. Bone and cartilage segmentation from multiplaner images using state of the art conovolutinal Neural Network, in: 13th Annual International Workshop on Osteoarthritis Imaging. IWOAI, Prince Edward Island.

Felson, D.T., Naimark, A., Anderson, J., Kazis, L., Castelli, W., Meenan, R.F., 1987. The prevalence of knee osteoarthritis in the elderly. the framingham osteoarthritis study. Arthritis Rheum. 30, 914–918. https://doi.org/10.1002/art.1780300811

Girshick, R., 2015. Fast R-CNN, in: Proceedings of the IEEE International Conference on Computer Vision. pp. 1440–1448. https://doi.org/10.1109/ICCV.2015.169

Goodfellow, I., Yoshua Bengio, Aaron Courville, 2016. deep learning .

He, K., Gkioxari, G., Dollár, P., Girshick, R., 2020. Mask R-CNN. IEEE Trans. Pattern Anal. Mach. Intell. 42, 386–397. https://doi.org/10.1109/TPAMI.2018.2844175

He, K., Gkioxari, G., Dollár, P., Girshick, R., 2017. Mask R-CNN. pp. 2961–2969.

He, K., Zhang, X., Ren, S., Sun, J., 2016. Deep Residual Learning for Image Recognition, in: 2016 IEEE Conference on Computer Vision and Pattern Recognition (CVPR). IEEE, pp. 770–778. https://doi.org/10.1109/CVPR.2016.90

Huang, G., Liu, Z., Van Der Maaten, L., Weinberger, K.Q., 2017. Densely connected convolutional networks, in: Proceedings - 30th IEEE Conference on Computer Vision and Pattern Recognition, CVPR 2017. pp. 2261–2269. https://doi.org/10.1109/CVPR.2017.243

Hunter, D.J., Guermazi, A., Lo, G.H., Grainger, A.J., Conaghan, P.G., Boudreau, R.M., Roemer, F.W., 2011. Evolution of semi-quantitative whole joint assessment of knee OA : MOAKS ( MRI Osteoarthritis Knee Score ) 19. https://doi.org/10.1016/j.joca.2011.05.004

Jaremko, J.L., Felfeliyan, B., Hareendranathan, A., Thejeel, B., Vanessa, Q.L., Østergaard, M., Conaghan, P.G., Lambert, R.G.W., Ronsky, J.L., Maksymowych, W.P., 2021. Volumetric quantitative measurement of hip effusions by manual versus automated artificial intelligence techniques: An OMERACT preliminary validation study. Semin. Arthritis Rheum. https://doi.org/10.1016/j.semarthrit.2021.03.009

Jaremko, Jacob L, Jeffery, D., Buller, M., Wichuk, S., McDougall, D., Lambert, R.G.W., Maksymowych, W.P., 2017. Preliminary validation of the Knee Inflammation MRI Scoring System (KIMRISS) for grading bone marrow lesions in osteoarthritis of the knee: Data from the Osteoarthritis Initiative. RMD Open 3, 1–9. https://doi.org/10.1136/rmdopen-2016-000355

Jaremko, Jacob L., Jeffery, D., Buller, M., Wichuk, S., McDougall, D., Lambert, R.G.W., Maksymowych, W.P., 2017. Preliminary validation of the Knee Inflammation MRI Scoring System (KIMRISS) for grading bone marrow lesions in osteoarthritis of the knee: Data from the Osteoarthritis Initiative. RMD Open 3, 1–9. https://doi.org/10.1136/rmdopen-2016-000355

John Ball; Maurice Rutherford Jeffrey; Jonas Henrik Kellgren;, 1963. The epidemiology of chronic rheumatism; Volume 2: Atlas of standard radiographs of arthritis. Oxford Blackwell Scientific Publications.

Kauffmann, C., Gravel, P., Godbout, B., Gravel, A., Beaudoin, G., Raynauld, J.P., Martel-Pelletier, J., Pelletier, J.P., De Guise, J.A., 2003. Computer-aided method for quantification of cartilage thickness and volume changes using MRI: Validation study using a synthetic model. IEEE Trans. Biomed. Eng. 50, 978–988. https://doi.org/10.1109/TBME.2003.814539

Lin, T.Y., Dollár, P., Girshick, R., He, K., Hariharan, B., Belongie, S., 2017. Feature pyramid

networks for object detection, in: Proceedings - 30th IEEE Conference on Computer Vision and Pattern Recognition, CVPR 2017. pp. 936–944. https://doi.org/10.1109/CVPR.2017.106

Lin, T.Y., Maire, M., Belongie, S., Hays, J., Perona, P., Ramanan, D., Dollár, P., Zitnick, C.L., 2014. Microsoft COCO: Common objects in context, in: Lecture Notes in Computer Science (Including Subseries Lecture Notes in Artificial Intelligence and Lecture Notes in Bioinformatics). Springer Verlag, pp. 740–755. https://doi.org/10.1007/978-3-319-10602-1_48

Liu, F., Zhou, Z., Jang, H., Samsonov, A., Zhao, G., Kijowski, R., 2017. Deep convolutional neural network and 3D deformable approach for tissue segmentation in musculoskeletal magnetic resonance imaging. Magn. Reson. Med. https://doi.org/10.1002/mrm.26841

Liu, F., Zhou, Z., Samsonov, A., Blankenbaker, D., Larison, W., Kanarek, A., Lian, K., Kambhampati, S., Kijowski, R., 2018. Deep learning approach for evaluating knee MR images: Achieving high diagnostic performance for cartilage lesion detection. Radiology 289, 160–169. https://doi.org/10.1148/radiol.2018172986

Otsu, N., 1979. A Threshold selection method from gray-level histograms. IEEE Trans. Syst. 62–66.

Perry, T.A., Gait, A., O'Neill, T.W., Parkes, M.J., Hodgson, R., Callaghan, M.J., Arden, N.K., Felson, D.T., Cootes, T.F., 2019. Measurement of synovial tissue volume in knee osteoarthritis using a semiautomated MRI-based quantitative approach. Magn. Reson. Med. 81, 3056–3064. https://doi.org/10.1002/mrm.27633

Peterfy, C.G., Guermazi, A., Zaim, S., Tirman, P.F.J., Miaux, Y., White, D., Kothari, M., Lu, Y., Fye, K., Zhao, S., Genant, H.K., 2004. Whole-organ magnetic resonance imaging score (WORMS) of the knee in osteoarthritis. Osteoarthr. Cartil. 12, 177–190. https://doi.org/10.1016/j.joca.2003.11.003

Prasoon, A., Petersen, K., Igel, C., Lauze, F., Dam, E., Nielsen, M., 2013. Deep feature learning for knee cartilage segmentation using a triplanar convolutional neural network, in: International Conference on Medical Image Computing and Computer-Assisted Intervention. Springer, pp. 246–253.

Quinn-Laurin, V., Bostick, G.P., Thejeel, B., Mandegaran, R., Steer, K.J.D., Lambert, R.G.W., Jaremko, J.L., 2021. Development of a technique for MRI gold-standard direct volumetric measurement of complex joint effusion, and validation at the hip. Skeletal Radiol. 50, 781–787. https://doi.org/10.1007/s00256-020-03630-6

Ren, S., He, K., Girshick, R., Sun, J., 2017. Faster R-CNN: Towards Real-Time Object Detection with Region Proposal Networks. IEEE Trans. Pattern Anal. Mach. Intell. 39, 1137–1149. https://doi.org/10.1109/TPAMI.2016.2577031

Roemer, F.W., Hunter, D.J., Winterstein, A., Li, L., Kim, Y.J., Cibere, J., Mamisch, T.C., Guermazi, A., 2011. Hip Osteoarthritis MRI Scoring System (HOAMS): Reliability and associations with radiographic and clinical findings. Osteoarthr. Cartil. 19, 946–962. https://doi.org/10.1016/j.joca.2011.04.003

Ronneberger, O., Fischer, P., Brox, T., 2015. U-net: Convolutional networks for biomedical image segmentation, in: Lecture Notes in Computer Science (Including Subseries Lecture Notes in Artificial Intelligence and Lecture Notes in Bioinformatics). Springer Verlag, pp. 234–241. https://doi.org/10.1007/978-3-319-24574-4_28

Salaffi, F., Carotti, M., Stancati, A., Grassi, W., 2005. Health-related quality of life in older adults with symptomatic hip and knee osteoarthritis: A comparison with matched healthy

controls. Aging Clin. Exp. Res. 17, 255–263. https://doi.org/10.1007/BF03324607

Shah, R.F., Martinez, A.M., Pedoia, V., Majumdar, S., Vail, T.P., Bini, S.A., 2019. Variation in the Thickness of Knee Cartilage. The Use of a Novel Machine Learning Algorithm for Cartilage Segmentation of Magnetic Resonance Images. J. Arthroplasty 34, 2210–2215. https://doi.org/10.1016/j.arth.2019.07.022

Sharif, B., Garner, R., Hennessy, D., Sanmartin, C., Flanagan, W.M., Marshall, D.A., 2017. Productivity costs of work loss associated with osteoarthritis in Canada from 2010 to 2031. Osteoarthr. Cartil. 25, 249–258. https://doi.org/10.1016/j.joca.2016.09.011

Unter Rote, G., 1991. COMPUTING THE MINIMUM HAUSDORFF DISTANCE BETWEEN TWO POINT SETS ON A LINE UNDER TRANSLATION. Inf. Process. Lett. 38, 123–127.

Wojna, Z., Ferrari, V., Guadarrama, S., Silberman, N., Chen, L.C., Fathi, A., Uijlings, J., 2019. The Devil is in the Decoder: Classification, Regression and GANs. Int. J. Comput. Vis. 127, 1694–1706. https://doi.org/10.1007/s11263-019-01170-8

Wright, E.A., Katz, J.N., Cisternas, M.G., Kessler, C.L., Wagenseller, A., Losina, E., 2010. Impact of knee osteoarthritis on health care resource utilization in a US population-based national sample. Med. Care 48, 785–791. https://doi.org/10.1097/MLR.0b013e3181e419b1

Zhang, R., Cheng, C., Zhao, X., Li, X., 2019. Multiscale Mask R-CNN–Based Lung Tumor Detection Using PET Imaging. Mol. Imaging 18. https://doi.org/10.1177/1536012119863531

Zhang, Y., Chu, J., Leng, L., Miao, J., 2020. Mask-refined R-CNN: A network for refining object details in instance segmentation. Sensors (Switzerland) 20. https://doi.org/10.3390/s20041010